\documentclass[aps,prd,reprint,showpacs,floatfix,longbibliography,nofootinbib,superscriptaddress]{revtex4-1}

\usepackage{tikz,graphicx,booktabs,multirow,array,microtype,mathtools,float}

\usepackage{centernot}

\graphicspath{{figures/}{figures3/}}
\usepackage{enumerate}   
\extrafloats{200}
\usepackage[colorlinks=true,backref=false, linktocpage=true,
citecolor=blue,
urlcolor=blue,linkcolor=blue,
pdfpagemode=UseOutlines]{hyperref}
\newcolumntype{C}{>{$}c<{$}}
\usepackage{dcolumn}

\usepackage{anyfontsize}
\usepackage[shortlabels]{enumitem}

\usepackage{amsfonts}
\AtBeginDocument{
\heavyrulewidth=.08em
\lightrulewidth=.05em
\cmidrulewidth=.03em
\belowrulesep=.65ex
\belowbottomsep=0pt
\aboverulesep=.4ex
\abovetopsep=0pt
\cmidrulesep=\doublerulesep
\cmidrulekern=.5em
\defaultaddspace=.5em
}
\def\fm {\mathop{\hbox{fm}}}

\def\Tr {\mathop{\hbox{Tr}}}

\usepackage{amsmath} 
\usepackage{dsfont}  
\usepackage{bm}      

\def\beq{\begin{equation}}
\def\eeq{\end{equation}}
\def\beqs#1\eeqs{\beq\begin{split} #1 \end{split}\eeq}

\long\def\comment#1{}

\def\norm#1{\left\| #1 \right\|}

\def\ket#1{\left| #1 \right\rangle}
\def\bra#1{\left\langle #1 \right|}

\def\opbraket#1#2#3{ \left\langle #1 \left| #2 \right| #3 \right\rangle}
\def\av#1{ \left\langle #1 \right\rangle }

\makeatletter
\renewcommand\onecolumngrid{%
  \do@columngrid{one}{\@ne}%
  \def\set@footnotewidth{\onecolumngrid}%
  \def\footnoterule{\kern-6pt\hrule width 1.5in\kern6pt}%
}
\renewcommand\twocolumngrid{%
  \def\footnoterule{\dimen@\skip\footins\divide\dimen@\thr@@
    \kern-\dimen@\hrule width.5in\kern\dimen@}%
  \do@columngrid{mlt}{\tw@}%
}
\makeatother

\begin{document}

\title{
Pion vector form factor from LapH-smeared currents in lattice QCD
}
%
\author{Frank~X.~Lee}
\email{fxlee@gwu.edu}
\affiliation{Physics Department, The George Washington University, Washington, DC 20052, USA}
\author{Andrei~Alexandru}
\email{aalexan@gwu.edu}
\affiliation{Physics Department, The George Washington University, Washington, DC 20052, USA}

\begin{abstract}
We explore a method to extract pion vector form factor from three-point functions using currents constructed from LapH-smeared quark fields.  LapH employs the low eigenmodes of the Laplacian Heaviside operator on the lattice to smear out the quark fields in spatial directions. The LapH-smeared quark propagator retains all symmetry properties of the original one and can be re-used to construct correlation functions at a reduced cost, with equal treatment of connected and disconnected diagrams. The calculations are performed on a two-flavor dynamical nHYP-smeared ensemble on the $24^3\times 48$ lattice at a spacing of 0.12 fm and pion mass of 315 MeV.
We show that an extrapolation in the number of low eigenmodes in the LapH subspace can reproduce the same form factor 
computed from the conventional SST method with point source, sink, and current. 
The study opens up the possibility of applying LapH-smeared currents to extract hadron form factors and polarizabilities including disconnected diagrams.  

\end{abstract}
\maketitle

\twocolumngrid
\section{Introduction}
\label{sec:intro}

The pion electromagnetic form factor, $F_\pi(Q^2)$, describes the electromagnetic structure of a charged pion and provides a basic test of the nonperturbative dynamics of quantum chromodynamics (QCD). It is defined through the matrix element of the electromagnetic current between pion states, with the normalization $F_\pi(0)=1$ fixed by charge conservation. The slope at $Q^2=0$ determines the pion mean-square charge radius. 

Lattice QCD provides a first-principles approach to the pion form factor through the evaluation of Euclidean correlation functions. Early calculations established the feasibility of extracting the electromagnetic matrix element from lattice three-point functions~\cite{WilcoxLiu:1987,Draper:1989,Bonnet:2005,Boyle:2008}. Subsequent studies employed different lattice actions, pion masses, and techniques for accessing momentum transfer~\cite{Brommel:2007,Brandt:2013,Frezzotti:2009,Aoki:2009pion,Nguyen:2011,Fukaya:2014,Feng_2020}. More recent calculations have extended these studies toward the physical pion mass and improved control of the form factor and charge radius~\cite{Koponen:2016,Alexandrou:2018pion,Gao:2021pion}. 

The conventional sequential-source technique (SST) is widely used to calculate connected three-point functions. With a fixed sink momentum and sink operator, the sequential propagator allows different current insertion times and momentum transfers to be evaluated without additional inversions. However, changing the sink setup generally requires new sequential inversions. The computational requirements become more demanding for quark-line disconnected contributions, which involve all-to-all propagators connecting many lattice sites. Stochastic methods and other all-to-all propagator techniques have been developed to address this problem~\cite{Dong:1994,Foley:2005,Bali:2010}.

An alternative is provided by the distillation method, in which quark fields are projected onto a subspace spanned by low eigenmodes of the spatial gauge-covariant Laplacian~\cite{Peardon:2009gh}. This construction is commonly referred to as Laplacian Heaviside (LapH) smearing. The resulting quark propagators are represented in a reduced subspace and can be combined with different hadron operators and current insertions. Distillation has been used in hadron spectroscopy, while stochastic LapH was developed to reduce the computational cost on larger lattices~\cite{Dudek:2010,Morningstar:2011ka}. 
A wavelet-based distillation method was developed recently~\cite{chakraborty2026}.

So far, LapH smearing has been mainly applied to hadron interpolating fields to improve their overlap with low-energy states. In the present work,  we investigate the use of LapH-smeared quark fields in the electromagnetic current itself. At finite LapH rank $N_v$, the smeared current differs from the conventional local current. The dependence on the number of retained eigenmodes must therefore be examined before the method can be used to extract physical matrix elements. 
In particular, we want to see if an extrapolation in the number of LapH eigenmodes can recover the form factor computed by the conventional SST method.
Such a comparison study has two potential benefits. 
One is to improve the signal-to-noise ratio, especially at large momentum transfers.
The other is to offer a viable alternative to improve the evaluation of disconnected diagrams.
The pion vector form factor provides a baseline test because its electromagnetic matrix element is the simplest.

Experimental measurements of the form factor have been carried out over a range of spacelike momentum transfers~\cite{Amendolia:1986,Volmer:2001,Horn:2006,Tadevosyan:2007}. The form factor has also been studied using perturbative QCD, chiral perturbation theory, and nonperturbative continuum approaches~\cite{Lepage:1980,Efremov:1980,Maris:2000,Gasser:1984,Gasser:1985,Bijnens:1998}.

The remainder of the paper is organized as follows. Section~\ref{sec:method} describes the three-point functions and the SST and LapH methods. The simulation results and comparison are presented in Sec.~\ref{sec:results}. The conclusions and outlook are given in Sec.~\ref{sec:con}. Raw ratios on which the results are based are collected in the appendix.

\onecolumngrid
\section{Methodology}
\label{sec:method} 

In the continuum, pion form factor $F_\pi (Q^2)$ is defined by the matrix element for the electromagnetic current of an on-shell pion which has the general Lorentz structure,
\beq
\langle \pi(p_f)|j_\mu| \pi(p_i) \rangle_{cont.} \equiv (p_f+p_i)_\mu F_\pi (Q^2),
\eeq
where $p_f$ is the four-momentum vector for the final pion and $p_i$ the initial pion.
We are interested in the space-like region ($Q^2>0$) defined by the four-momentum transfer squared in Minkowski spacetime,
\beq
Q^2\equiv - (p_f-p_i)^2=-(E_f-E_i)^2+ \bm q^2,
\eeq
where $E_i = \sqrt{\bm p_i^2 +m_\pi^2}$ and
$E_f = \sqrt{\bm p_f^2 +m_\pi^2}$ are the energies, and $\bm q \equiv \bm p_f -\bm p_i$ the three-momentum transfer.
The normalization convention for the form factor is $F_\pi(0)=1$.

\begin{figure}[h!]
\includegraphics[scale=0.5]{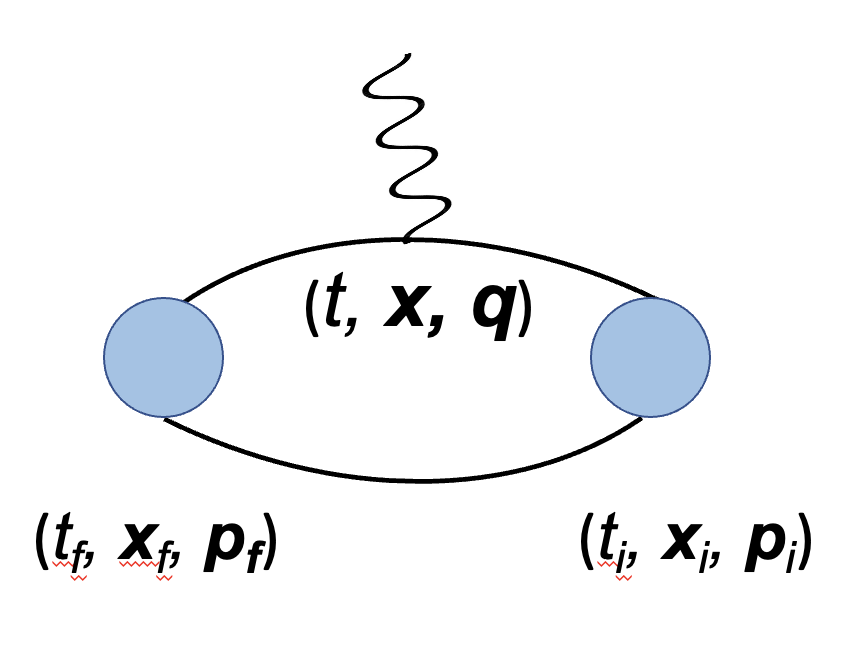}
\caption{Depiction of a three-point function. A pion is created at point $(t_i,\bm x_i)$ and annihilated at  $(t_f,\bm x_f)$, with an electromagnetic current coupling to it at $(t,\bm x)$ in between. The pion carries initial energy $E_i=\sqrt{\bm p_i^2+m^2}$ and momentum $\bm p_i$, and final $E_f=\sqrt{\bm p_f^2+m^2}$ and $\bm p_f$, with momentum transfer $\bm q= \bm p_f- \bm p_i $ in the process.}
\label{3pt-pic}
\end{figure}

On the lattice, the matrix element is normalized by a kinematical factor relative to the continuum,
\beq
\langle \pi (p_f) |j_\mu | \pi (p_i) \rangle_{lat.} \equiv 
{\langle \pi (p_f) |j_\mu | \pi (p_i)  \rangle_{cont.} \over 2\sqrt{E_f E_i} }. 
\eeq
Momentum-projected interpolators each carry a factor $V^{-1/2}$ ($V=N_x N_y N_z$),
\beq
\psi_\pi(t,\bm p) = V^{-1/2}\sum_{\bm x} e^{-i\bm p\cdot\bm x}\psi_\pi(t,\bm x),
\eeq
so that the states are normalized as continuum-like plane waves in a finite box.
The current is a density and is \emph{not} scaled by $V^{-1/2}$;
its Fourier transform is $\sum_{\bm x} e^{+i\bm q\cdot\bm x} j_4(t,\bm x)$.
At $\bm q=0$ this is the charge operator, so the ratio of conserved-current three- and two-point functions approaches the charge.
The lattice matrix element in momentum space is the Euclidean time-ordered correlator of these operators (with $\mu=4$),
\beqs
G_{3}(t_f,t,t_i,\bm p_f,\bm p_i) &= 
\langle \Omega|{\cal T}\psi_\pi(t_f,\bm p_f) \Big(\sum_{\bm x} e^{+i\bm q\cdot\bm x} j_4(t,\bm x)\Big) \psi_\pi^\dagger(t_i,\bm p_i) |\Omega\rangle \\
&= 
 V^{-1}
 \sum_{\bm x_f,\bm x,\bm x_i} e^{-i\bm p_f\cdot (\bm x_f-\bm x) } e^{-i\bm p_i\cdot (\bm x-\bm x_i) } 
 \langle \Omega|{\cal T}\psi_\pi(t_f,\bm x_f) j_4(t,\bm x) \psi_\pi^\dagger(t_i,\bm x_i) |\Omega\rangle,
 \label{3pt_lat}
\eeqs
where $\Omega$ denotes the QCD vacuum, $\bm q=\bm p_f-\bm p_i$, and the Fourier phase on the current is $e^{+i\bm q\cdot\bm x}$.
The three-point function is illustrated in Fig.~\ref{3pt-pic}

The three-point function is often analyzed in conjunction with a two-point function between the initial point and an open point,
\beqs
G_2(t,t_i,\bm p)= 
\langle \Omega|\psi_\pi(t,\bm p) \psi_\pi^\dagger(t_i,\bm p) |\Omega\rangle
=
V^{-1}
\sum_{\bm x,\bm x_i} e^{-i\bm p\cdot (\bm x-\bm x_i) } \langle \Omega|\psi_\pi(t,\bm x) \psi_\pi^\dagger(t_i,\bm x_i) |\Omega\rangle,
 \label{2pt_lat}
\eeqs
projected to momentum $\bm p$ of choice.

At the hadronic level,
the physics content of the three-point function can be revealed by inserting two complete sets of intermediate states ($|n\rangle$ with $\bm p_f$ and $|n'\rangle$ with $\bm p_i$), and assuming periodic boundary condition in time (which involves wrap-around effects associated with the time extent of the lattice $T$),
\beqs
G_3(t_f,t,t_i,\bm p_f,\bm p_i) &=
\sum_{n,n'} 
\langle\Omega|\psi_\pi|n\rangle \langle n|j_4|n'\rangle \langle n'|\psi^\dagger_\pi|\Omega\rangle
\big(e^{-E_n (t_f-t) } e^{-E_{n'}(t-t_i) } 
+ e^{-E_n (T-t_f+t) } e^{-E_{n'} (T-t+t_i) }\big) 
\\& \hspace{-17mm}\xrightarrow{t_f-t, t-t_i \gg 1} 
\langle\Omega|\psi_\pi|\pi\rangle \langle \pi|j_4|\pi'\rangle \langle \Omega|\psi_\pi|\pi'\rangle^*
\big(e^{-E_\pi (t_f-t) } e^{-E'_\pi (t-t_i) } 
+ e^{-E_\pi (T-t+t_f) } e^{-E'_\pi(T-t+t_i) }\big)
\\&
+\langle\Omega|\psi_\pi|\pi\rangle \langle \pi|j_4|1'\rangle \langle \Omega|\psi_\pi|1'\rangle^*
\big(e^{-E_\pi (t_f-t) } e^{-E'_1 (t-t_i) } 
+ e^{-E_\pi (T-t+t_f) } e^{-E'_1 (T-t+t_i) }\big)
\\&
+\langle\Omega|\psi_\pi|1\rangle \langle 1|j_4|\pi'\rangle \langle \Omega|\psi_\pi|\pi'\rangle^*
\big(e^{-E_1 (t_f-t) } e^{-E'_\pi (t-t_i) } 
+ e^{-E_1 (T-t+t_f) } e^{-E'_\pi (T-t+t_i) }\big)
\\&
+\langle\Omega|\psi_\pi|1\rangle \langle 1|j_4|1'\rangle \langle \Omega|\psi_\pi|1'\rangle^*
\big(e^{-E_1 (t_f-t) } e^{-E'_1 (t-t_i) } 
+ e^{-E_1 (T-t+t_f) } e^{-E'_1 (T-t+t_i) }\big)+\cdots.
\label{3pt_hadron}
\eeqs
The leading contributions at large time separations are also displayed if we retain the two lowest states (ground state $|\pi\rangle$ and first excited state $|1\rangle$).
The various $\langle\Omega|\psi_\pi|\cdots\rangle $ are vacuum-to-pion overlaps. The form factor of interest $F_\pi(Q^2)$ is contained in the matrix element $\langle \pi|j_4|\pi'\rangle$. The $\langle 1|j_4|1'\rangle$ transition matrix element is relatively small and can be usually ignored, but the $\langle \pi|j_4|1'\rangle$ and $\langle 1|j_4|\pi'\rangle$ transitions can play a role if precision is desired. 

Similarly, the 
two-point function is given by inserting one complete set of states, 
\beqs
G_2(t,t_i,\bm p)&=\sum_{n}
|\langle\Omega|\psi_\pi|n\rangle|^2   
\big(e^{- E_n (t-t_i)}+e^{- E_n (T-t+t_i) }\big) \\&
\xrightarrow{t-t_i \gg 1} 
|\langle\Omega|\psi_\pi|\pi\rangle|^2
\big(e^{- E_\pi (t-t_i)}+e^{- E_\pi (T-t+t_i) }\big)
+|\langle\Omega|\psi_\pi|1\rangle|^2
\big(e^{- E_1 (t-t_i)}+e^{- E_1 (T-t+t_i) }\big),
\label{2pt_hadron}
\eeqs
noting that the time dependence in this function always comes as the difference $t-t_i$.

At the lowest order, the form factor $F_\pi(Q^2)$ can be isolated from the  ratio of three-point to two-point functions, after accounting for normalization factors,
\beqs
R_1(t_f,t,t_i,\bm p_f,\bm p_i) \equiv 
{G_3(t_f,t,t_i,\bm p_f,\bm p_i)\over G_2(t_f-t_i,\bm p_f)}
\to 
F_\pi(Q^2)
{E_i+E_f  \over 2\sqrt{E_iE_f}} 
\sqrt{C(\bm p_i) \over C(\bm p_f)  } 
e^{- (E_i-E_f)(t-t_i) },
\eeqs
where time dependence on $t_f$ is eliminated.
The energies $E_i(\bm p_i)$ and $E_f(\bm p_f)$,
and the overlaps $C(\bm p_i)=|\langle\Omega|\psi_\pi|\pi(\bm p_i)\rangle|^2 $
and $C(\bm p_f)=|\langle\Omega|\psi_\pi|\pi(\bm p_f)\rangle|^2 $, can be determined from the two-point functions in Eq.\eqref{2pt_hadron} once kinematics for $\bm p_i$ and $\bm p_f$ are specified.
Alternatively,  $F_\pi(Q^2)$  can be isolated directly from the ratio,
\beqs
R_2(t_f,t,t_i,\bm p_f,\bm p_i) \equiv 
{2\sqrt{E_iE_f} \over E_i+E_f  } 
{G_3(t_f,t,t_i,\bm p_f,\bm p_i)\over G_2(t_f-t_i,\bm p_f)}
\sqrt{G_2(t_f-t,\bm p_i) G_2(t-t_i,\bm p_f) G_2(t_f-t_i,\bm p_f) 
\over G_2(t_f-t,\bm p_f) G_2(t-t_i,\bm p_i) G_2(t_f-t_i,\bm p_i)}
\to  
F_\pi(Q^2),
\label{eq:R2}
\eeqs
where the leading time dependence and ground-state overlaps are canceled out. We emphasize that the ratios are defined in the limit of large time separations ($t_f-t, t-t_i \gg 1$), and are subject to wrap-around and excited-state effects on the finite lattice. 
The interpolators in $G_3$ and $G_2$ carry the same $V^{-1/2}$ factors, while the current is unnormalized, so $G_3/G_2$ has no leftover volume dependence and the ratio retains the form above.

At the quark level, 
vacuum expectation values of operators ${\cal O}$ are evaluated by path integrals in Euclidean spacetime for the QCD action consisting of gauge and fermion parts $S_{QCD} = S_G + S_F$,
\beq
\langle\Omega|{\cal O}[U,\psi,\bar{\psi}] |\Omega\rangle \equiv 
{ \int {\cal D}U {\cal D}\psi  {\cal D}\bar{\psi} {\cal O}[U,\psi,\bar{\psi}] e^{-S_{QCD}}
\over 
\int {\cal D}U {\cal D}\psi  {\cal D}\bar{\psi} e^{-S_{QCD}} }
= { \int {\cal D}U {\cal O}(M^{-1})\text{det}M e^{-S_G}
\over 
\int {\cal D}U \text{det}M  e^{-S_G } }.
\eeq
In the second step, the fermion degrees of freedom are integrated out, assuming $S_F=\bar{\psi} M(\{U\}) \psi$ with $M$ representing the quark matrix (Dirac operator).
The result is a path integral over gauge links over a product of inverses whose specific form depends on the operator ${\cal O}$. The integral is computed numerically by Monte Carlo methods, treating $\text{det}M  e^{-S_G}$ as a weighting factor. A single quark propagator $M^{-1}$ is obtained by inverting the quark matrix on the generated Monte Carlo configurations. 
Expectation values are simply statistical averages of correlation functions composed of quark propagators over the configurations.

In the case of a charged pion ($\pi^+$), we use the following interpolating field to excite it from the vacuum,
\beq
\psi_\pi(x)=\bar{d}(x)\gamma_5 u(x).
\eeq
For the lattice version of $j_4$,
we consider two options. One is a local current defined by,
\beqs
j^{(\text{local})}_4(x) &\equiv Z_V \left[q_u (2\kappa_u) \bar{u}(x)\gamma_4 u(x) + q_d (2\kappa_d)\bar{d}(x)\gamma_4 d(x) \right]. \\
\label{eq:j1PC}
\eeqs
The factor $2\kappa$ comes from the action for Wilson-type fermions. 
The charge factors are $q_u=2/3$ and $q_d=-1/3$ in units of electric charge $|e|$.
The advantage of this current operator is that it leads to simple correlation functions. 
The drawback is one must determine the renormalization constant $Z_V$ on the lattice. 
The other option is the conserved current (or point-split current), 
\beqs
 j^{(\text{conserved})}_4 (x) \equiv 
 q_u {\kappa_u} \big[ 
-\bar{u}(x) (1-\gamma_4) U_4(x) u(x+\hat{4}) 
+
\bar{u}(x+\hat{4}) (1+\gamma_4 ) U_4^\dagger(x) u(x) 
\big] 
+ (u \leftrightarrow d).
\label{eq:j4con}
\eeqs
This operator has $Z_V\equiv 1$ on the lattice by construction, but leads to more complicated correlation functions due to the explicit presence of gauge fields $U_\mu(x)$. 

The two-point function in Eq.\eqref{2pt_lat}
can be expressed in terms of quark propagators by contracting out all quark-antiquark pairs (Wick contraction),
\beq
G_2(t,t_i,\bm p)= 
V^{-1}
\sum_{\bm x,\bm x_i} e^{-i\bm p\cdot (\bm x-\bm x_i) } \langle 
\text{tr}\left[S_{u}(t_i,\bm x_i; t,\bm x)\gamma_5S_{u}(t,\bm x; t_i,\bm x_i)\gamma_5 \right]
+ \text{tr}\left[ S_{d}(t_i,\bm x_i; t,\bm x)\gamma_5S_{d}(t,\bm x; t_i,\bm x_i)\gamma_5\right] \rangle,
\label{2pt}
\eeq
where $\langle \cdots \rangle$ denotes the configuration average.
 The trace is over spin and color. 
 In the isospin limit ($\kappa_u=\kappa_d$), the two terms are equal.
The $S(t,\bm x;t_i, \bm x_i)$ denotes a fully-interacting quark propagator from point $(t_i,\bm x_i)$ to point $(t,\bm x)$ in $12 \times 12$  spin-color space on the lattice, obtained as the inverse of the quark matrix $M$.  

\begin{figure}[h]
\includegraphics[scale=0.5]{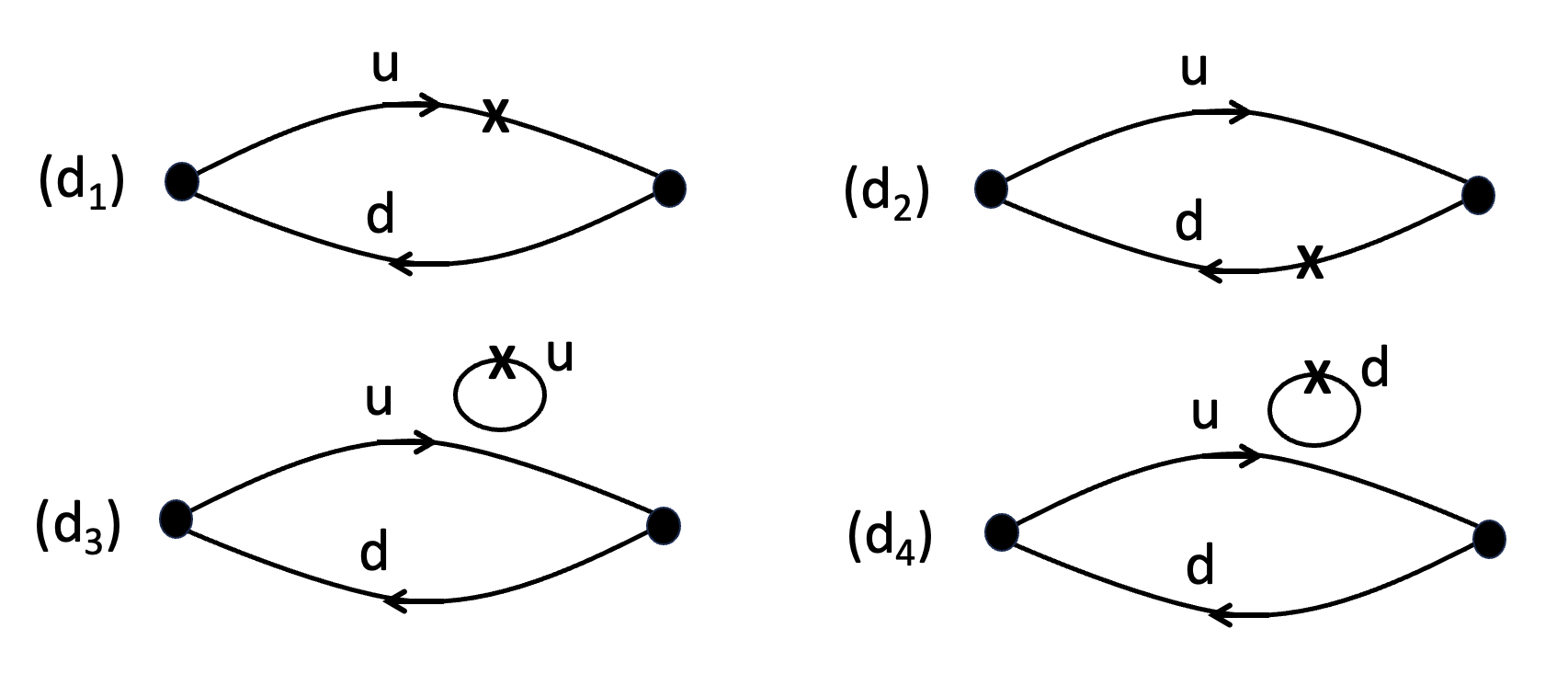}
\caption{Quark-line diagrams from Wick contraction of a three-point function for pion form factor.
Diagrams ($d_1$) and ($d_2$) are connected insertions.
Diagrams ($d_3$) and ($d_4$)  are disconnected insertions.}
\label{fig:3pt}
\end{figure}

Wick contraction of the three-point function in Eq.\eqref{3pt_lat}, using the local current in Eq.\eqref{eq:j1PC}, is given by,
\beqs
G_3(t_f,t,t_i,\bm p_f,\bm p_i) 
=\sum_k G^{(k)}_3(t_f,t,t_i,\bm p_f,\bm p_i)
=\sum_k  Z_V V^{-1} \sum_{\bm x_f,\bm x,\bm x_i} e^{-i\bm p_f\cdot (\bm x_f-\bm x) } e^{-i\bm p_i\cdot (\bm x-\bm x_i)}
\langle 
 d_k
 \rangle
 ,
   \label{eq:3pt}
\eeqs
where 
\beqs
d_1 &=-q_u (2\kappa_u) \text{tr}\left[S_u(t,\bm x; t_f,\bm x_f)\gamma_5 S_d(t_f,\bm x_f; t_i, \bm x_i)\gamma_5 S_u(t_i,\bm x_i; t,\bm x)\gamma_4\right],\\
d_2 &=-q_d (2\kappa_d) \text{tr}\left[S_u(t_i,\bm x_i; t_f,\bm x_f)\gamma_5 S_d(t_f,\bm x_f; t,\bm x)\gamma_4 S_d(t,\bm x; t_i,\bm x_i)\gamma_5 \right],\\
d_3 &=q_u (2\kappa_u)\text{tr}\left[S_u(t_i,\bm x_i; t_f,\bm x_f)\gamma_5 S_d(t_f,\bm x_f; t_i,\bm x_i)\gamma_5\right] \text{tr}\left[S_u(t,\bm x; t,\bm x)\gamma_4 \right],\\
d_4 &=q_d (2\kappa_d)\text{tr}\left[S_u(t_i,\bm x_i; t_f,\bm x_f)\gamma_5 S_d(t_f,\bm x_f; t_i,\bm x_i)\gamma_5\right] \text{tr}\left[S_d(t,\bm x; t,\bm x)\gamma_4 \right].
\label{3pt_wick}
\eeqs
The $d_1$ and $d_2$ terms in this equation are connected insertions (CI), 
and $d_3$ and $d_4$ disconnected insertions (DI).
 In the isospin limit, $d_1$ and $d_2$ are equal aside from the charge factors ($q_u$ and $q_d$) in the CI. The same is true for $d_3$ and $d_4$ in the DI.
They can be represented  by distinct quark-line diagrams depicted in Fig.~\ref{fig:3pt}.

Before evaluating the three-point function in Eq.\eqref{eq:3pt}, we examine its symmetry properties~\cite{Draper:1989}.
Under charge conjugation, the relation $C\gamma_\mu C^{-1}= -\gamma_\mu^T$ holds in any representation. It leads to the fact that the quark matrix satisfies, on a configuration-by-configuration basis, 
$C M(\{U\}) C^{-1}= M(\{U^*\})^T$, so does its inverse 
$C M^{-1}(\{U\}) C^{-1}= M^{-1}(\{U^*\})^T$. 
Here the transpose is over color, spin, and space-time indices.
It also implies that configuration $\{U\}$ and its complex conjugate $\{U^*\}$ have equal weight in the path integral,
$\text{det} M(\{U^*\}) e^{-S_G(U^*)} = \text{det} M(\{U\}) e^{-S_G(U)}$. 
So the three-point function (without the momentum projections) can be expressed as the average of the two distributions, 
\beq
\langle d(\{U\})\rangle  = {1\over 2} \langle d(\{U\}) + d(\{U^*\}) \rangle.
\label{green}
\eeq
The disconnected contribution (see $d_3$ and $d_4$ in Eq.\eqref{3pt_wick}) is the product of two traces. One of them is essentially the pion correlator $\text{tr}[SS^\dagger]$ which is invariant under charge conjugation. The other trace $\text{tr}[S\gamma_4]$ changes sign due to $\gamma_4$. Consequently, the DI changes sign when the configuration is complex conjugated, $d_{3,4}(\{U\}) = - d_{3,4}(\{U^*\})$. It follows immediately from Eq.\eqref{green} that the disconnected contribution $d_{3,4}(\{U\})$ vanishes in the ensemble average.
On the other hand, the connected contribution ($d_1$ and $d_2$) is a single trace in the form of $\text{tr}[S\gamma_5 S\gamma_5 S\gamma_4]$. Under another charge conjugation relation 
$(C\gamma_5) \gamma_\mu (C\gamma_5)^{-1}= +\gamma_\mu^*$, 
the propagator obeys
$(C\gamma_5) M^{-1}(\{U\}) (C\gamma_5)^{-1}= M^{-1}(\{U^*\})^*$
which means the CI acquires complex conjugation but not sign, $d_{1,2}(\{U\}) = d_{1,2}(\{U^*\})^*$.
Then it follows from Eq.\eqref{green} that the connected contribution $d_{1,2}(\{U\})$ is real-valued in the ensemble average. The imaginary part in $G_3$ is due entirely to momentum factors $e^{i\bm p\cdot\bm x}$.

Additionally, since the accessible momenta are quantized on the periodic spatial volume, the discrete rotational symmetry of the ensemble average means that one may further reduce noise by averaging over permutations of the components of the three-momentum transfer $\bm q$.

\subsection{SST method} 
The evaluation of the connected insertions (CI) in Eq.\eqref{3pt_wick} starts by taking advantage of translation invariance to eliminate one of the spatial sums ($\bm x_i$) and placing the source at a chosen point $(\bm x_{\mathrm{src}}, t_{\mathrm{src}})$.
Each interpolator contributes $V^{-1/2}$ and dropping $\sum_{\bm x_i}$ supplies one $V$, so neither the two- nor the three-point function carries a leftover volume ($V=N_x N_y N_z$).
For the remaining part, a brute-force evaluation requires three quark propagators per configuration; two of them must be repeated when the current insertion point and momentum transfer vary. The computation can be greatly reduced by a method called the SST (Sequential Source Technique). In the isospin limit, diagram $d_1$ with the local current can be expressed in matrix form (suppressing spin and color indices),
\beq
G^{(1),\text{loc}}_3(t_f,t,t_{\mathrm{src}},\bm p_f,\bm p_i) = 
- Z_V (2\kappa) q_u  \sum_{\bm x} 
e^{+i\bm q\cdot(\bm x-\bm x_{\mathrm{src}})}
\langle 
\text{tr}\left[S_{t_f,\bm p_f}(t,\bm x) 
\,\gamma_5 S_u^\dagger(t,\bm x;t_{\mathrm{src}},\bm x_{\mathrm{src}})\gamma_5\,
\gamma_4 \right]
 \rangle.
 \label{3pt_sst} 
\eeq
This three-point function is visually represented by the $d_1$ diagram in Fig.~\ref{fig:3pt}.
The third quark line of $d_1$ runs from the insertion back to the source via $\gamma_5$-hermiticity,
$S_u(t_{\mathrm{src}},\bm x_{\mathrm{src}};t,\bm x)=\gamma_5 S_u^\dagger(t,\bm x;t_{\mathrm{src}},\bm x_{\mathrm{src}})\gamma_5$.
Two sets of $12$ linear equations of the form $Mv=b$ have to be solved per source point.
One is to calculate $S_u(t,\bm x;t_{\mathrm{src}},\bm x_{\mathrm{src}})$ from the source to any other point on the lattice.
The same quark propagator is used to evaluate the two-point function in Eq.\eqref{2pt} in the isospin limit,
\beq
G_2(t,t_{\mathrm{src}},\bm p)= 
2  \sum_{\bm x} e^{-i\bm p\cdot (\bm x-\bm x_{\mathrm{src}}) } \langle 
\Tr\left[S_{u}(t,\bm x; t_{\mathrm{src}},\bm x_{\mathrm{src}})S_{u}^\dagger(t,\bm x; t_{\mathrm{src}},\bm x_{\mathrm{src}})\right ]
\rangle.
\label{2ptsst}
\eeq
%
The two interpolators each contribute $V^{-1/2}$, which cancels the volume from translation invariance, so no extra $V$ remains.
The other is a {\em sequential} quark propagator, $S_{t_f,\bm p_f}$, that solves 
\beq
[M  S_{t_f,\bm p_f}](t,\bm x) = 
\delta(t-t_f)
e^{-i\bm p_f\cdot(\bm x-\bm x_{\mathrm{src}})}\gamma_5 S_u(t_f,\bm x; t_{\mathrm{src}},\bm x_{\mathrm{src}})\gamma_5.
\label{eq:sst_seq}
\eeq
The sign of $\bm p_f$ in the sequential source matches the sink Fourier factor $e^{-i\bm p_f\cdot\bm x_f}$ in Eq.\eqref{3pt_lat}.
Note that the current insertion is left open so different momentum transfer and different current (scalar, tensor, conserved, etc) can be considered post-inversion.
This is the so-called ``fixed ends, open insertion" SST approach. Alternatively, if the focus is on a given insertion (both $\bm q$ and current type) but different hadrons,  then one can re-organize the calculation in the ``open ends, fixed insertion" SST approach.

The same sequential propagator gives the conserved (point-split) insertion of Eq.\eqref{eq:j4con} with no extra inversions,
\beqs
&G^{(1),\mathrm{cons}}_3(t_f,t,t_{\mathrm{src}},\bm p_f,\bm p_i)
=
- q_u \kappa
\sum_{\bm x}
e^{+i\bm q\cdot(\bm x-\bm x_{\mathrm{src}})}
\\
&
\Big\langle
\text{tr}\big[
S_{t_f,\bm p_f}(t,\bm x)\,\gamma_5 S_u^\dagger(t+1,\bm x;t_{\mathrm{src}},\bm x_{\mathrm{src}})\gamma_5\,(1+\gamma_4)U_4^\dagger(x)
-S_{t_f,\bm p_f}(t+1,\bm x)\,\gamma_5 S_u^\dagger(t,\bm x;t_{\mathrm{src}},\bm x_{\mathrm{src}})\gamma_5\,(1-\gamma_4)U_4(x)
\big]
\Big\rangle,
\label{3pt_sst_cons}
\eeqs
where we used the fact that $Z_V\equiv 1$ for the conserved current, both hop terms of $j_4(x)$ share the phase $e^{+i\bm q\cdot(\bm x-\bm x_{\mathrm{src}})}$, and $x+\hat 4$ is the timeslice $t+1$ at the same spatial site.

\subsection{LapH method} 

Here we explore an alternative way to evaluate the three-point functions in Eq.\eqref{eq:3pt}, namely, the LapH method originally proposed for spectrum calculations in Refs~\cite{Morningstar:2013,Dudek:2012,Bulava:2016,Lang:2025}.
It can be understood as a form of smearing of all the quark fields in the system (source, sink, and current),
with the added bonus that the calculation can be completely expressed in terms of a quark-propagator
reduced to a subspace of slowly moving quark states. The smearing is purely in the spatial direction and
it is gauge covariant by construction~\cite{Morningstar:2004}. As such the smeared quark fields have the same transformation
properties under lattice symmetry transformations.  Interpolators built out of smeared fields
have the same quantum numbers as the ones built using the original fields.

The smearing is constructed using the eigenvectors of the three-dimensional covariant LapH  operator on a given time slice $t$,
\beq
\Delta_t = -\sum_{k=1}^3 \nabla_k(t)\nabla_k(t)^\dagger\,,
\eeq
with components in spatial and color space,
\beqs
\Delta_t^{ab}(\bm x,\bm y)& =
\sum^3_{k=1}[U_k^{ab}(\bm x,t) \delta(\bm x+\hat{k},\bm y) 
      +U_k^{ba}(\bm y,t)^*\delta(\bm x-\hat{k},\bm y)-2\delta^{ab}\delta(\bm x,\bm y)] \,.
\eeqs
We label the eigenvectors by $|s;t\rangle$ whose eigensystem equation is, 
\beq 
\sum_{\bm y,c}\Delta_t^{ac}(\bm x,\bm y) \phi_{st}(\bm y,c) 
= \lambda_{s}(t) \phi_{st}(\bm x,a),
\eeq
with eigenvector  $\phi_{st}(\bm x,c) = \langle \bm x,c |s;t\rangle$.

The operator is defined as negative-definite with its eigenvalues all negative. We sort the eigenvalues in ascending order (not by absolute value). We define the LapH smearing operator:
\beq
\mathcal{P}(t) \equiv \sum_{s=1}^{N_v} \ket{s;t} \bra{s;t}\,,
\eeq
which is the projector on the space spanned by the $N_v$ lowest frequency eigenmodes of the Laplacian operator.
The smearing operator only acts on the spatial and color space. A LapH-smeared quark field is simply,
$ 
\widetilde{q}(t) \equiv \mathcal{P}(t) q(t).
$

The action of the smearing operator $\mathcal{P}$ can be illustrated by acting on a point source. The magnitude of 
$\norm{(\mathcal{P}\delta_{\bm x})(\bm y)}$ decays like a gaussian away from the source, $\exp(-\norm{\bm x-\bm y}^2/(2r^2))$. 
The smearing radius $r$ depends on the energy cutoff, $\Lambda=-\lambda_{N_v}(t)$. 
We can determine the optimal $\Lambda$ by tuning individual operators to minimize the errorbars of the effective mass at a
fixed time. For the ensemble in this study, we tune the maximum number of Laplacian eigenvectors to $N_v=100$ which corresponds to $r\approx 0.5\fm$~\cite{Guo:2016}. 

Bilinears $\widetilde{\bar{q}}\Gamma\widetilde{q}$ have the same transformation properties as $\bar{q}\Gamma q$ and we can use them as building blocks to construct correlation functions.  
The advantage of this substitution is that, on one hand, the correlation functions will be less noisy,
since the overlap of these interpolators with the physical states is better when we choose $N_v$ appropriately.
On the other hand, the calculation of all the correlation functions (connected and disconnected) requires only the evaluation
of the quark propagators from $4\times N_v\times N_t$ sources.
A regular quark propagator in the correlation functions is replaced with a smeared version,
\beq
\av{q(t)\bar{q}(t')} \rightarrow \av{\widetilde{q}(t)\widetilde{\bar{q}}(t')}
=\mathcal{P}(t)M^{-1}(t,t')\mathcal{P}(t')\equiv \tilde{M}^{-1}   \,.
\label{eq:mtilde}
\eeq

The projection to the $4 N_v$-dimensional LapH subspace uses the same Wick-contracted correlation functions as in the SST method.
For example, the $d_1$ term of the three-point function in Eq.\eqref{eq:3pt} is replaced with a LapH-smeared version,
\beqs
& G^{(1)}_3(t_f,t,t_i,\bm p_f,\bm p_i) \\
 &= -Z_V (2\kappa) q_u  
\sum_{\bm x_f,\bm x,\bm x_i}
\langle \Tr\Big[
\gamma_4 e^{i\bm (p_f-p_i)\cdot \bm x} {S}_u(t,\bm x; t_f,\bm x_f)
\gamma_5 e^{-i\bm p_f\cdot \bm x_f}{S}_d(t_f,\bm x_f; t_i, \bm x_i)
\gamma_5 e^{i\bm p_i\cdot \bm x_i} {S}_u(t_i,\bm x_i; t,\bm x)\Big]
 \rangle \\
&\to -Z_V (2\kappa) q_u  \langle \Tr\Big[ 
\tilde{\Gamma}_4(t,\bm p_f-\bm p_i) \widetilde{S}_u(t, t_f) \tilde{\Gamma}_5(t_f,-\bm p_f) \widetilde{S}_d(t_f,t_i) \tilde{\Gamma}_5(t_i,\bm p_i)\widetilde{S}_u(t_i, t)
 \Big] \rangle,
 \label{eq:CI}
\eeqs
where the spatial sums are absorbed in the matrix notation. The trace acts on matrices of dimension $4N_v$.
The $\widetilde{\Gamma}$ are $(4 N_v)\times(4 N_v)$ matrices carrying the gamma matrix structure and momentum factors associated with the three time points,
\beqs
\tilde{\Gamma}_\mu(t,\bm p)_{s,s'}^{\alpha\beta} &=
\opbraket{s;t}{\gamma_\mu^{\alpha\beta} e^{+i\bm p\cdot \bm x}}{s';t} \,,\\
\eeqs
where $s, s'$ are eigenvector indices and $\alpha, \beta$ are spinorial indices. The $\widetilde{S}$ are shorthand for LapH-smeared quark propagators in Eq.\eqref{eq:mtilde} which take the generic form that highlights its time dependence,
\beqs
\widetilde{S}(t,t')_{s,s'}^{\alpha\beta} &\equiv \opbraket{s;t}{S^{\alpha\beta}}{s';t'}.
\eeqs
We see that the three-point function in the LapH method is a straightforward evaluation of the Wick contractions, taking the product of three terms of the same form $\tilde{\Gamma}\widetilde{S}$ at the three time points $(t_f, t, t_i)$. No sequential source is needed.

In the same notation, the two-point function in Eq.\eqref{2ptsst} is replaced by,
\beqs
G_2(t,t_{\mathrm{src}},\bm p) &= 
2  \sum_{\bm x} e^{-i\bm p\cdot (\bm x-\bm x_{\mathrm{src}}) } \langle 
\Tr\Big[S_{u}(t,\bm x; t_{\mathrm{src}},\bm x_{\mathrm{src}})S_{u}^\dagger(t,\bm x; t_{\mathrm{src}},\bm x_{\mathrm{src}})\Big ] \rangle\\
&\to 2  \langle \Tr\Big[ 
 \tilde{\Gamma}_5(t,-\bm p) \widetilde{S}_u(t, t_{src}) 
\tilde{\Gamma}_5(t_{src},\bm p)\widetilde{S}_u(t_{src}, t)
 \Big] \rangle,
 \label{g2laph}
 \eeqs
%

Since LapH-smeared quark propagators are essentially all-to-all propagators; the disconnected insertion is on equal footing with the connected insertion.
For example, the DI in Eq.\eqref{eq:3pt} ($d_3$ term) is simply translated into,
\beqs
& G^{(3)}_3(t_f,t,t_i,\bm p_f,\bm p_i) \\
 &= Z_V (2\kappa) q_u  
\sum_{\bm x_f,\bm x,\bm x_i}
\langle \Tr \bigg[ 
\Tr\Big[\gamma_4 e^{i\bm (p_f-p_i)\cdot \bm x} {S}_u(t,\bm x; t,\bm x)\Big]
\gamma_5 e^{-i\bm p_f\cdot \bm x_f}{S}_d(t_f,\bm x_f; t_i, \bm x_i)
\gamma_5 e^{i\bm p_i\cdot \bm x_i} {S}_u(t_i,\bm x_i; t,\bm x_f)\bigg]
 \rangle \\
&\to Z_V (2\kappa) q_u  \langle \Tr\bigg[ 
\Tr\Big[\tilde{\Gamma}_4(t,\bm p_f-\bm p_i) \widetilde{S}_u(t, t)\Big]
\tilde{\Gamma}_5(t_f,-\bm p_f) \widetilde{S}_d(t_f,t_i) 
\tilde{\Gamma}_5(t_i,\bm p_i)\widetilde{S}_u(t_i, t_f)
 \bigg] \rangle,
 \label{eq:DI}
\eeqs
which is written to mirror the form in the CI in Eq.\eqref{eq:CI}. The $\widetilde{\Gamma}$ factors are the same. Only time points in $\widetilde{S}(t,t')$ and trace placement are different.
The equal treatment means that the computational cost is about the same between connected and disconnected diagrams in the LapH method because they share the same set of LapH-smeared quark propagators.
The savings could be significant for disconnected diagrams.
For example, the evaluation of LapH-smeared $\widetilde{S}$ only requires $4 N_v\times N_t$ inversions compared to $N=N_x\times N_y\times N_z\times N_t\times 12$ inversions for an all-to-all propagator. On the lattice $N=24^3\times 48$ and $N_v=100$ used in this study, it leads to a reduction factor of $24^3\times 12 / 400 \approx 415$.
Additional savings can come from factorization of spinorial and LapH subspaces in tensor products~\cite{Guo:2016}.

Additional comments about the LapH method are in order. 
Although LapH smearing in Eq.\eqref{eq:mtilde} incurs an additional cost than a conventional inversion for $M^{-1}$,
an advantage of the method is that we can separate the 
calculation of the smeared quark propagator $\tilde{M}^{-1}$ (also known as {\it perambulators}) from computing the correlation functions. 
Once the perambulators are computed, they become building blocks that can be re-used to construct different correlation functions in different projects, without having to redo the inversions. 
Another point is that the smeared interpolating fields have the same symmetry properties as the un-smeared versions even when the number of Laplacian eigenvectors $N_v$ is small. In the two-point function, $N_v$ controls the smearing radius and the overlap with the physical states.  In the three-point function, $N_v$ must be taken to infinity to recover the physical coupling of the current to the quarks.

\section{Simulation results and discussion}
\label{sec:results}
\onecolumngrid

In Table~\ref{tab:ens}, we list the parameters for the two-flavor dynamical nHYP clover action~\cite{nhyp2001,nhyp2009,nhyp2020,Sheikholeslami:1985,Hasenfratz:2007} used in this study.
\begin{table}[ht]
\centering
\caption{Ensemble and analysis parameters used in the study.
}
\label{tab:ens}
\small
\begin{tabular}{lc}
\toprule
Quantity & Value \\
\midrule
Volume $N_x^3\times N_t$ & $24^3\times 48$ \\
$\beta$, $\kappa$, $c_{\mathrm{sw}}$ & $7.1$, $0.1282$, $1$ \\
nHYP smearing $(\alpha_1,\alpha_2,\alpha_3)$ & $(0.75,\,0.6,\,0.3)$ \\
Lattice spacing $a$ & $0.121\,\mathrm{fm}$ \\
$a^{-1}$ & $1.63\,\mathrm{GeV}$ \\
Pion mass  & $am_\pi=0.1933(5)$ $\to$ $315(1)\,\mathrm{MeV}$ \\
Configurations generated & $300$  \\
Jackknife bin & $4$ \\
LapH rank $N_v$ & $\{70,80,90,100\}$ \\
\bottomrule
\end{tabular}
\end{table}

Kinematics are common to SST and LapH:
final pion at rest $\mathbf{p}_f=\mathbf{0}$, initial pion moving with momentum $\mathbf{p}_i=-\mathbf{q}$.
We consider 4 momentum shells for the momentum transfer,
\begin{equation}
\mathbf{q}={2\pi\over aN_x}\mathbf{n}, \text{ with }
  \mathbf{n}=(0,0,0), (1,0,0), (1,1,0), (1,1,1),
\end{equation}
also in shorthand as $q0$, $q1$, $q11$, $q111$, respectively.  Rotationally equivalent $\mathbf{q}$ in a shell are averaged.

We aim to compare form factors from the SST and LapH methods at three nonzero momenta. We then fit the form factors with a single-pole VMD model to access the charge radius.

\subsection{Two-point function}

\begin{figure}[ht]
\centering
\includegraphics[width=0.6\textwidth]{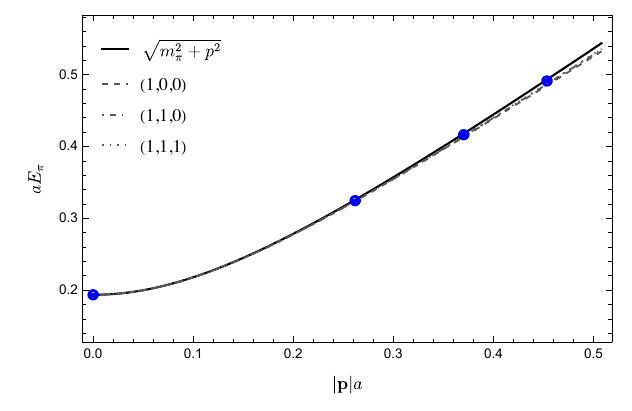}
\caption{Dispersion relation: lattice points compared with the continuum dispersion
$\sqrt{m_\pi^2+\mathbf{p}^2}$ and the lattice free-boson
relations along the $(1,0,0)$, $(1,1,0)$, and $(1,1,1)$
directions.}
\label{fig:disp}
\end{figure}

Pion energies $E_\pi(\mathbf{q})$ for shared rest-frame and moving-frame states are extracted from LapH-smeared ($N_v=100$) two-point functions in Eq.\eqref{g2laph}, using two-cosh fits. 
They are given in Table~\ref{tab:energies} and are used for all kinematic factors.
The dispersion relation in lattice units is plotted in Fig.~\ref{fig:disp}.
\begin{table}[ht]
\centering
\caption{Pion energies from two-point functions,
used for all kinematical factors.}
\label{tab:energies}
\small
\begin{tabular}{lccc}
\toprule
Shell & $\mathbf{n}$ & $aE_\pi$ & $E_\pi$ [MeV] \\
\midrule
$q0$ & $(0,0,0)$ & $0.1933(5)$ & $315(1)$ \\
$q1$ & $(1,0,0)$ & $0.3243(8)$ & $529(1)$ \\
$q11$ & $(1,1,0)$ & $0.4161(11)$ & $679(2)$ \\
$q111$ & $(1,1,1)$ & $0.4910(12)$ & $801(2)$ \\
\bottomrule
\end{tabular}
\end{table}

\subsection{SST results }
\label{sec:SST}

For SST inversions, six point sources at $\mathbf{x}=0$ and
$t_{\mathrm{src}}=0,8,16,24,32,40$ are averaged.
The sequential inversion is at the pion sink with
$\mathbf{p}_f=\mathbf{0}$. 
Sink-source separations are $t_f\in\{10,12,14,16\}$.
Per configuration this is $6\times(12+4\times 12)=360$ inversions
(forward plus sequential).

The connected insertion of the three-point functions ($d_1+d_2$) is evaluated by SST using the local current $j_4$ and point
interpolators (sequential source at $\mathbf{p}_f=\mathbf{0}$).
Shell-averaged $G_3$, $G_2(\mathbf{p}_f)$, $G_2(\mathbf{p}_i)$
enter the ratio in Eq.\eqref{eq:R2}.
Jackknife covariances are built with bin size~$4$.

All ratios are wrap-corrected. 
With periodic time,
$G_2(\tau)=G_2^{\mathrm{fwd}}(\tau)\,w(E,\tau)$ 
where $G_2^{\mathrm{fwd}}$ is the forward-propagating part of the correlator (see Eq.\eqref{2pt_hadron}), and $w$ the wrap factor given by,
\begin{equation}
  w(E,\tau)=1+e^{-E(N_t-2\tau)}.
\label{eq:w}
\end{equation}
The measured ratio is then $R_2^{\mathrm{meas}}=R_2^{\mathrm{fwd}}/W$ with
\begin{equation}
  W
  =\sqrt{\frac{w(E_i,t)\,w(E_i,t_f)\,w(E_f,t_f-t)\,w(E_f,t_f)}
  {w(E_i,t_f-t)\,w(E_f,t)}}.
\end{equation}
The wrap correction applied to the mid-ratio ($t=t_f/2$) 
 uses the sink-time factor $t_f$ only,
\beqs
  w_0(t_f)=w(E_f,t_f),\quad
  w_q(t_f)=\sqrt{w(E_i,t_f)\,w(E_f,t_f)}.
\label{eq:wfac}
\eeqs
The masses $E_f=E_\pi(\mathbf{0})$ and $E_i=E_\pi(\mathbf{q})$
entering~\eqref{eq:wfac} are the shared values of
Table~\ref{tab:energies}. 
The wrap-corrected midpoint combination is
\begin{equation}
  R_2^{\mathrm{fwd}}(t_f,\mathbf{q})
  =\,w_q(t_f)\,R_2(t_f/2,t_f,\mathbf{q}),
\end{equation}
and at $\mathbf{q}=\mathbf{0}$,
$w_q\to w_0=1+e^{-m_\pi(N_t-2t_f)}$.
The wrap-around has a non-negligible impact on the ensemble in this study. Without this factor the $\mathbf{q}=0$ midpoints fall with $t_f$;
after correction the four $t_f$ collapse onto a common plateau,
see Fig.~\ref{fig:sst-r2} in Appendix~\ref{app:r2}.

For the form factor, we use the estimator at $t_f=16$, normalized by the ratio at $\bm q=0$,
\begin{equation}
  F_\pi(Q^2)
  =\frac{w_q(t_f)\,R_2(t_f/2,t_f,\mathbf{q})}
        {w_0(t_f)\,R_2(t_f/2,t_f,\mathbf{0})}
  =\frac{R_2^{\mathrm{fwd}}(t_f,\mathbf{q})}
        {R_2^{\mathrm{fwd}}(t_f,\mathbf{0})},
\label{eq:F}
\end{equation}
so $F_\pi(0)=1$ identically.
The renormalization constant $Z_V$ for the local current is separately extracted  from,
\beq
Z_V={1\over R_2^{\mathrm{fwd}}(t_f,\bm 0)},
\label{eq:zv}
\eeq
under the assumption of current conservation on the lattice, which is checked independently using the conserved current in Eq.\eqref{eq:j4con}.

We fit the form factor to the simple  vector meson dominance (VMD) model,
\begin{equation}
  F_{\mathrm{VMD}}(Q^2)=\frac{1}{1+Q^2/M^2}.
\label{eq:VMD}
\end{equation}
with a single parameter $M$, using the full jackknife covariance of
the $F_\pi$ points.
The mean-square charge radius is  
$
\langle r^2\rangle={6/ M^2}.
$

The results for $F_\pi(Q^2)$, $Z_V$, and VMD  are given in the summary Table~\ref{tab:F-fitg2} under  \texttt{SST measured}.
The $q111$ point suffers a large relative error (about $\sim 40\%$ ) and
does not constrain the fit tightly; $\chi^2/\mathrm{dof}$ is
consequently very small.
The large error is traced to a single two-point function in the ratio.
At the midpoint, Eq.\eqref{eq:R2} contains the source-momentum
two-point function at the full separation,
\begin{equation}
  R_2(t_f/2,t_f,\mathbf{q})\;\propto\;
  \frac{G_3(t_f,t_f/2,\mathbf{q})}
  {\sqrt{G_2(t_f,\mathbf{p}_f)\,G_2(t_f,\mathbf{p}_i)}}.
\end{equation}
The signal-to-noise ratio of $G_2(t_f,\mathbf{p}_i)$ decays roughly as
$e^{-(E_\pi(\mathbf{p}_i)-m_\pi)t_f}$, which has relative values of $44, 25, 9.5,1.1$ for q0, q1, q11, q111 at $t_f=16$.
In other words, the correlator $G_2(t_f=16,\bm q_i=q111)$ is consistent with zero at the $1\sigma$ level, 
and it sits in a square root in the denominator of $R_2$.
Furthermore, the q111 shell has a single momentum, so there is no rotational
average to reduce the noise.
Holding $G_2(t_f,\mathbf{p}_i)$ at its ensemble mean in every jackknife
sample, and leaving everything else as measured, reduces the error of
$F_\pi(q111)$ at $t_f=16$ by more than a factor of 6.
Almost all of the error therefore comes from this one number.

To enhance the signal for the $q111$ point, we replace the noisy factor by a fit to the full $G_2(t,\mathbf{p}_i)$,
 which is constrained by the precise small-$t$ data.
For each shell, the shell-averaged SST $G_2(t,\mathbf{p}_i)$ is fit
  with a two-state cosh,
  \begin{equation}
    G_2(t)=A_0\bigl(e^{-E_0t}+e^{-E_0(N_t-t)}\bigr)
          +A_1\bigl(e^{-E_1t}+e^{-E_1(N_t-t)}\bigr),
  \label{eq:g2fit}
  \end{equation}
  over $t=4$--$18$ with the full correlated covariance.
  The ground-state energy $E_0$ is fixed to the value
  $E_\pi(\mathbf{q})$ of Table~\ref{tab:energies}; $A_0$, $A_1$, and
  $E_1$ are free.
Only $G_2(t_f,\mathbf{p}_i)$ in Eq.\eqref{eq:R2} is replaced by
  the fitted value.
The  $G_3$, the other five $G_2$ factors, the prefactor, and the wrap
  factor~\eqref{eq:wfac} are unchanged, and $F_\pi$ follows
  Eq.\eqref{eq:F} at $t_f=16$.

Table~\ref{tab:g2fit} lists the central fits.
For $q1$ and $q11$ the fitted $G_2(16)$ agrees with the measured one
within $1\sigma$.
For $q111$ the fit lies about $0.5\sigma$ above the measured value.
Since $F_\pi\propto G_2(t_f,\mathbf{p}_i)^{-1/2}$ this lowers the midpoint.
The fitted $G_2(16)$ changes by at most $3\%$ at small times $t_{\min}=3$--$5$.
The results for $F_\pi(Q^2)$,  $Z_V$, and VMD after the fitted $G_2$ correction are given in the summary Table~\ref{tab:F-fitg2} under  \texttt{SST fitted}.
\begin{table*}
\centering
\caption{Two-state fits from Eq.\eqref{eq:g2fit} to the two-point function
$G_2(t,\mathbf{p}_i)$ used in the SST ratio with $E_0$ fixed and $t=4$--$18$.
The last two columns compare the measured and fitted $G_2$ at $t=16$.
Their difference at $q111$ is the source of the large uncertainty observed in the raw SST result.}
\label{tab:g2fit}
\small
\begin{tabular}{lcccccc}
\toprule
Shell & $aE_0$ (fixed) & $\chi^2/\mathrm{dof}$ & $A_0$ & $aE_1$ &
  $G_2(16)$ measured & $G_2(16)$ fit \\
\midrule
$q1$ & $0.3243$ & $1.91$ & $0.930(18)$ & $1.09(7)$ &
  $5.36(21)\times10^{-3}$ & $5.21\times10^{-3}$ \\
$q11$ & $0.4161$ & $1.06$ & $0.749(19)$ & $1.20(7)$ &
  $9.2(1.0)\times10^{-4}$ & $9.63\times10^{-4}$ \\
$q111$ & $0.4910$ & $2.58$ & $0.655(29)$ & $1.24(12)$ &
  $1.8(1.6)\times10^{-4}$ & $2.54\times10^{-4}$ \\
\bottomrule
\end{tabular}
\end{table*}

\subsection{LapH results}
\label{sec:loc}

We consider two prescriptions that change different parts of the projected operator 
in order to check possible rank dependence in the results.

In the \texttt{LapH loc} prescription,
for each $N_v\in\{70,80,90,100\}$, shell-averaged three-point
functions are built from \texttt{loc} data at that $N_v$, and
two-point functions use the same $N_v$.
$R_2$ follows Eq.\eqref{eq:R2} with energies from Table~\ref{tab:energies}.
Raw and wrap-corrected $R_2(t)$ for all shells and $N_v$ are
given in Figs.~\ref{fig:loc-r2-a}--\ref{fig:loc-r2-b} in Appendix~\ref{app:r2} for visual inspection.

At $t_f=16$ the wrap-corrected ratio for $F_\pi(Q^2)$ is evaluated
for each $N_v$ and each momentum shell.
A linear model,
\begin{equation}
  R(N_v)=A+{B\over N_v}
\label{eq:linNv}
\end{equation}
is fit by ordinary least squares on the four $N_v$ points;
the intercept $A$ and its error (and the $\{A,B\}$ covariance)
come from a jackknife over configurations
(cross-$N_v$ correlations included).
The extrapolated $A$ is identified with $F_\pi(Q^2)$.
A joint jackknife over the three nonzero shells supplies the
$3\times 3$ covariance used in the VMD fit.

\begin{figure*}[ht]
\centering
\includegraphics[width=0.85\textwidth]{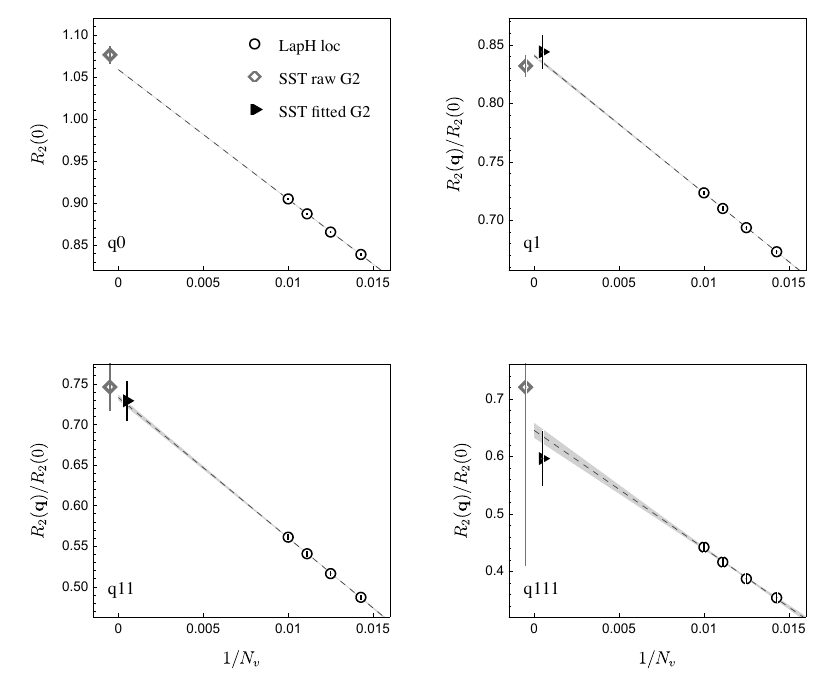}
\caption{
Linear extrapolation of LapH \texttt{loc} ratios in $1/N_v$
at $q0,q1,q11,q111$.
Open circles: LapH jackknife means and errors.
Grey band and dashed line: linear fit~\eqref{eq:linNv} with jackknife $\{A,B\}$ covariance.
At $1/N_v=0$, the two SST points (offset horizontally for visibility) with measured
$G_2$ (grey diamond) and fitted $G_2(t_f,\mathbf{p}_i)$
(black triangle) are plotted for comparison.
The measured-$G_2$ SST error at $q111$ extends beyond the
$y$-range.}
\label{fig:fitg2-RNv}
\end{figure*}

Table~\ref{tab:R-loc} gives the mid-ratios versus $N_v$ and the
extrapolated intercepts. 
Fig.~\ref{fig:fitg2-RNv} shows $R$ versus $1/N_v$ with the
jackknife error band and comparison with the SST points at $1/N_v=0$.
The results for $F_\pi(Q^2)$,  $Z_V$,  and VMD  are given in the summary Table~\ref{tab:F-fitg2} under \texttt{LapH loc}.

\begin{table*}[ht]
\centering
\caption{LapH \texttt{loc} and \texttt{loc\_cur} ratios as a function of $N_v$.
The $q0$ row is $R_2^{\mathrm{w}}(0)$ which yields $Z_V$ by Eq.\eqref{eq:zv}.
The $q1,q11,q111$ rows are normalized to the $R_2^{\mathrm{w}}(0)$, see Eq.\eqref{eq:F}. 
The last column is the linear intercept $A$ in Eq.\eqref{eq:linNv}.
}
\label{tab:R-loc}
\begin{tabular}{lccccc}
\toprule
\texttt{loc} & $N_v=70$ & $80$ & $90$ & $100$ & $A$ ($1/N_v\to0$) \\
\midrule
 $q0$  & 0.8390(6) & 0.8656(5) & 0.8873(5) & 0.9050(5) & 1.0586(5)\\
 $q1$ & $0.6730(11)$ & $0.6936(11)$ & $0.7100(11)$ & $0.7234(11)$
  & $0.8408(14)$  \\
 $q11$ & $0.4869(24)$ & $0.5161(24)$ & $0.5406(25)$ & $0.5610(25)$
  & $0.7329(34)$ \\
 $q111$ & $0.354(8)$ & $0.387(8)$ & $0.416(8)$ & $0.442(8)$
  & $0.645(13)$  \\
  \midrule
\texttt{loc\_cur} & $N_v=70$ & $80$ & $90$ & $100$ & $A$ ($1/N_v\to0$) \\
 \midrule
$q0$  & 0.8389(6) & 0.8656(5) & 0.8873(5) & 0.9050(5) & 1.0589(5)\\
 $q1$ & $0.6731(11)$ & $0.6936(11)$ & $0.7100(11)$ & $0.7234(11)$
  & $0.8405(13)$  \\
 $q11$ & $0.4869(22)$ & $0.5160(23)$ & $0.5405(24)$ & $0.5610(25)$
  & $0.7329(33)$  \\
 $q111$ & $0.358(7)$ & $0.390(7)$ & $0.418(8)$ & $0.442(8)$
  & $0.637(12)$  \\
\bottomrule
\end{tabular}
\end{table*}

In the \texttt{LapH loc\_cur} prescription, only the current rank varies,
while every two-point function in~\eqref{eq:R2} is taken at
$N_v=100$.
Prefactors and wrap masses still use Table~\ref{tab:energies}.
Mid-ratios, linear $1/N_v$ extrapolation~\eqref{eq:linNv}, joint
covariance, and VMD are repeated.
The ratio grids are given in 
Figs.~\ref{fig:loccur-r2-a}--\ref{fig:loccur-r2-b} in Appendix~\ref{app:r2}.

Numerically \texttt{loc\_cur} tracks \texttt{loc} closely.
The large $\chi^2/\mathrm{dof}\sim 30$ again signals a poor
single-pole description of the LapH $Q^2$ shape.
VMD with $F(0)=1$ fixed is a one-parameter description and
  is not expected to work at large $Q^2$ on this ensemble.
The results for $F_\pi(Q^2)$, $Z_V$,  and VMD  are given in the summary Table~\ref{tab:F-fitg2} under  \texttt{LapH loc\_cur}.

\subsection{Comparison of SST and LapH}
\label{sec:summary}

\begin{figure*}[ht]
\centering
\includegraphics[width=0.9\textwidth]{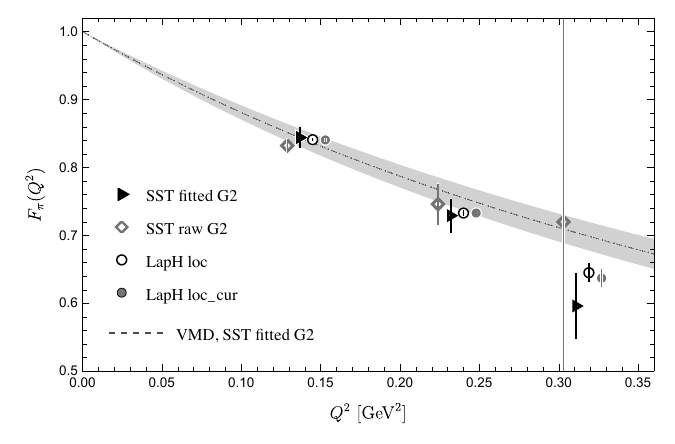}
\caption{Comparison of $F_\pi(Q^2)$ between SST and LapH methods.
The four points at the same $Q^2$ are offset horizontally for visibility.
Dashed curve with a grey band is the VMD fit to the fitted-$G_2$ SST values.
The measured-$G_2$ SST error at $(1,1,1)$ extends beyond the $y$-range.}
\label{fig:fitg2-f}
\end{figure*}

Table~\ref{tab:F-fitg2} compares the two SST methods (with measured and fitted
$G_2(t_f,\mathbf{p}_i)$) to the two LapH methods (\texttt{loc} and \texttt{loc\_cur} ).
Fig.~\ref{fig:fitg2-f} is a visual representation of Table~\ref{tab:F-fitg2} for the form factor.
LapH \texttt{loc} and LapH \texttt{loc\_cur} agree with each other;
both give a slightly larger $M$ (smaller radius) than SST and a
much larger VMD $\chi^2$, reflecting tension between the three
precise LapH $F_\pi(Q^2)$ points and a single pole.
SST and LapH agree well at $q1$ and
$q11$; the SST $q111$ point is not competitive.

The $Z_V$ for the ensemble in this study is around 0.94.
For reference, a previous determination~\cite{hoffmann2007} found $Z_V\sim 0.97$ using the same action, but at a finer lattice spacing than the one used here, see Table~\ref{tab:ens}.
In any event, it indicates the improved properties of the nHYP clover action over the standard Wilson action (with $Z_V$ around 0.6 for similar lattice spacing).
There is still a slight tension in $Z_V$ between the two methods that warrants further study. Possible causes include the excited state effects in the three-point functions.
The results for $F_\pi(Q^2)$ at nonzero $Q^2$ seem less affected possibly because the effects are attenuated in the ratios in  Eq.\eqref{eq:F}.

The fitted-SST for VMD fit is poorer than for the measured $G_2$
($\chi^2/\mathrm{dof}=3.7$ against $0.006$), mainly because the $q111$
point now has a small enough error to lie about $2.3\sigma$ below the single-pole curve.
The LapH VMD fits have even larger $\chi^2/\mathrm{dof}$ because of more precise $F_\pi(Q^2)$ values, and tension with a single-pole, single-parameter VMD. Better determination may require more sophisticated models such as the $z$-expansion~\cite{zexpand2015,Gao:2021pion} to cover the whole range of $Q^2$.

\begin{table*}[ht]
\centering
\caption{Comparison of $F_\pi(Q^2)$, $Z_V$, and VMD between SST and LapH
methods.
}
\label{tab:F-fitg2}
\small
\begin{tabular}{lcccccc}
\toprule
& && \multicolumn{2}{c}{SST, $G_2(t_f,\mathbf{p}_i)$} &
  \multicolumn{2}{c}{LapH, $N_v\to\infty$} \\
\cmidrule(lr){4-5}\cmidrule(lr){6-7}
&  $\mathbf{n}$ & $Q^2$ [GeV$^2$]  &
  measured & fitted & \texttt{loc} & \texttt{loc\_cur}  \\
\midrule
$Z_V$  & $(0,0,0)$ & 0 & $0.929(9)$ & $0.929(9)$ &
  $0.9446(4)$ & $0.9444(4)$ \\
$F_\pi$ & $(1,0,0)$ & $0.137$  & $0.832(9)$ & $0.844(14)$ &
  $0.8408(14)$ & $0.8405(13)$ \\
$F_\pi$ & $(1,1,0)$ & $0.232$  & $0.746(29)$ & $0.729(24)$ &
  $0.7329(34)$ & $0.7329(33)$ \\
$F_\pi$ & $(1,1,1)$ & $0.311$  & $0.72(31)$ & $0.596(47)$ &
  $0.645(13)$ & $0.637(12)$ \\
\midrule
\multicolumn{7}{l}{VMD, SST measured $G_2$:
   $M=0.824(26)\,\mathrm{GeV}$,
  $\chi^2/\mathrm{dof}=0.006$,
  $\langle r^2\rangle=0.344(22)\,\mathrm{fm}^2$} \\
  \multicolumn{7}{l}{VMD, SST fitted $G_2$:
   $M=0.860(43)\,\mathrm{GeV}$,
  $\chi^2/\mathrm{dof}=3.7$,
  $\langle r^2\rangle=0.316(32)\,\mathrm{fm}^2$} \\
\multicolumn{7}{l}{VMD, \texttt{LapH loc}:
   $M=0.844(4)\,\mathrm{GeV}$,
  $\chi^2/\mathrm{dof}=28.7$,
  $\langle r^2\rangle=0.328(3)\,\mathrm{fm}^2$} \\
\multicolumn{7}{l}{VMD, \texttt{LapH loc\_cur}:
   $M=0.842(4)\,\mathrm{GeV}$,
  $\chi^2/\mathrm{dof}=32.6$,
  $\langle r^2\rangle=0.329(3)\,\mathrm{fm}^2$} \\
\bottomrule
\end{tabular}
\end{table*}

\clearpage
\onecolumngrid
\section{Conclusion and Outlook}
\label{sec:con}

We have calculated the pion vector form factor using LapH-smeared currents and compared the results with those obtained from the conventional SST method. As far as we are aware, this is the first study of form factors using smeared currents.

The calculation was performed on a two-flavor nHYP-smeared ensemble at $m_\pi\simeq315$ MeV. Four LapH ranks, $N_v=70$, 80, 90, and 100, were used to examine the dependence of the form factor on the number of retained Laplacian eigenmodes. We found that a linear extrapolation in $1/N_v$ provides a good estimator for the form factor and $Z_V$ in the large-$N_v$ limit. The robustness of the agreement was tested with two different SST analyses and two different LapH constructions. The main results are summarized in Table~\ref{tab:F-fitg2} and Fig.~\ref{fig:fitg2-f}. The wrap-around effects in the two-point function are important in achieving the agreement on the ensemble used in this study. 

Overall, the LapH results have much smaller statistical uncertainties than the SST results. The primary reason is that the low LapH eigenmodes do not introduce additional noise beyond the gauge noise, unlike stochastic LapH~\cite{Morningstar:2011ka}. 
At large momentum transfers the SST method suffers a rapid loss of signal on the given ensemble. The large SST uncertainty at the highest momentum considered is traced mainly to the moving-pion two-point function at the full source-sink separation. Replacing this noisy factor by a two-state fit can reduce the uncertainty significantly, in addition to more statistics.  This improvement does not constitute a complete removal of excited-state effects from the three-point function. Such effects require additional analysis of the source-sink separation and transition matrix elements, as discussed in lattice studies of excited-state contamination~\cite{excited2025,Capitani:2012,Mastropas:2014}.

We also performed one-pole vector-meson-dominance (VMD) fits to the form factor as a secondary check. 
The fits have large $\chi^2/\mathrm{dof}$ values, indicating that the single-pole form does not describe the three precise LapH points satisfactorily. The charge radii inferred from these fits should therefore be regarded as model-dependent estimates rather than precision determinations.

The present study is exploratory in nature. Although careful attention has been paid to the statistical uncertainty and correlations in the data, 
several systematic uncertainties remain in the calculation. The linear extrapolation in $1/N_v$ has been tested using only four LapH ranks, and the quoted uncertainties do not include a systematic error from the extrapolation form. 
Additional ranks would help establish its reliability. 
Calculations at different lattice spacings and pion masses would be required for a controlled physical charge-radius determination.

Computationally, the LapH method by itself is not necessarily less costly than the SST method because of the overhead to compute LapH-smeared quark propagators (perambulators). However, the LapH method has a decisive advantage if the perambulators are pre-computed (only once) and re-used as building blocks to construct correlation functions in different projects. 
More savings can be achieved when the matrix multiplications in the correlation functions become as costly as the inversions, due to the reduction into the LapH $4N_v\times 4N_v$ subspace. 

An important next step is  to establish the LapH method for disconnected diagrams which are expensive to compute and often suffer from large statistical uncertainties in conventional methods. A major advantage of the LapH method is that it 
can greatly reduce the cost of disconnected diagrams while maintaining good signal-to-noise ratio. 
The connected and disconnected diagrams share the same LapH-smeared propagators and cost about the same to compute.
We were unable to check this claim because the disconnected insertions vanish in the vector channel. However, in the scalar channel disconnected insertion has been shown to be significant~\cite{G_lpers_2014}.
A comparison study in the scalar channel between the conventional treatment of disconnected diagrams and the LapH method proposed here would be very interesting. 

 We ultimately aim to use LapH currents in correlation functions with multiple electromagnetic insertions, such as the four-point functions used to determine hadron polarizabilities in lattice QCD~\cite{Wilcox:2021,Lee:2023,LeeMagnetic:2023,LeeNeutral:2024,Wang:2024,Feng:2022rkr}. 
In fact, four-point functions were the original motivation that led to the present study of three-point functions.

\begin{acknowledgments}
This work was supported in part by U.S. Department of Energy under  Grant~No.~DE-FG02-95ER40907. A combination of local clusters and Perlmutter at DOE-sponsored NERSC are used to carry out the calculations. 
Assistance is acknowledged from GW graduate student Hanxing Jiang, and CalTech undergraduate Jack Sullivan. 

\end{acknowledgments}

\clearpage
\onecolumngrid
\appendix
\section{Ratios of three-point to two-point functions}
\label{app:r2}

Here we document the insertion-time dependence of the
three-point-function ratios Eq.\eqref{eq:R2} used in the study. 
They are plotted as a function of $t-{t_f\over 2}$, centered around the midpoint between pion source and sink.
The figures serve as diagnostics of the underlying correlators and of
the stated wrap correction. 
They show both the raw
$R_2$ estimates and the values after applying the finite-temporal-extent
wrap correction of Eq.\eqref{eq:wfac}. Keeping both versions visible
makes it possible to distinguish the correction for backward propagation
around the periodic temporal boundary from any residual dependence on
insertion time or source-sink separation.

The four momentum shells are ordered as $\mathbf n=(0,0,0)$,
$(1,0,0)$, $(1,1,0)$, and $(1,1,1)$, with the same conventions
used in the main text. Markers identify the source-sink separation $t_f=10,12,14,16$. Final results for $F_\pi$  are taken from the mid-point of the largest source-sink separation $t_f=16$.
The $\mathbf n=(0,0,0)$ channel is particularly useful for examining
the expected charge normalization and $Z_V$. At nonzero momentum, any remaining
separation dependence after wrap correction must not automatically be
interpreted as a finite-temperature effect: excited states and statistical
fluctuations can also contribute. No additional excited-state subtraction
is performed on these displayed midpoint ratios.

\begin{figure}[ht]
\centering
\includegraphics[width=0.98\textwidth,height=0.6\textheight,keepaspectratio]{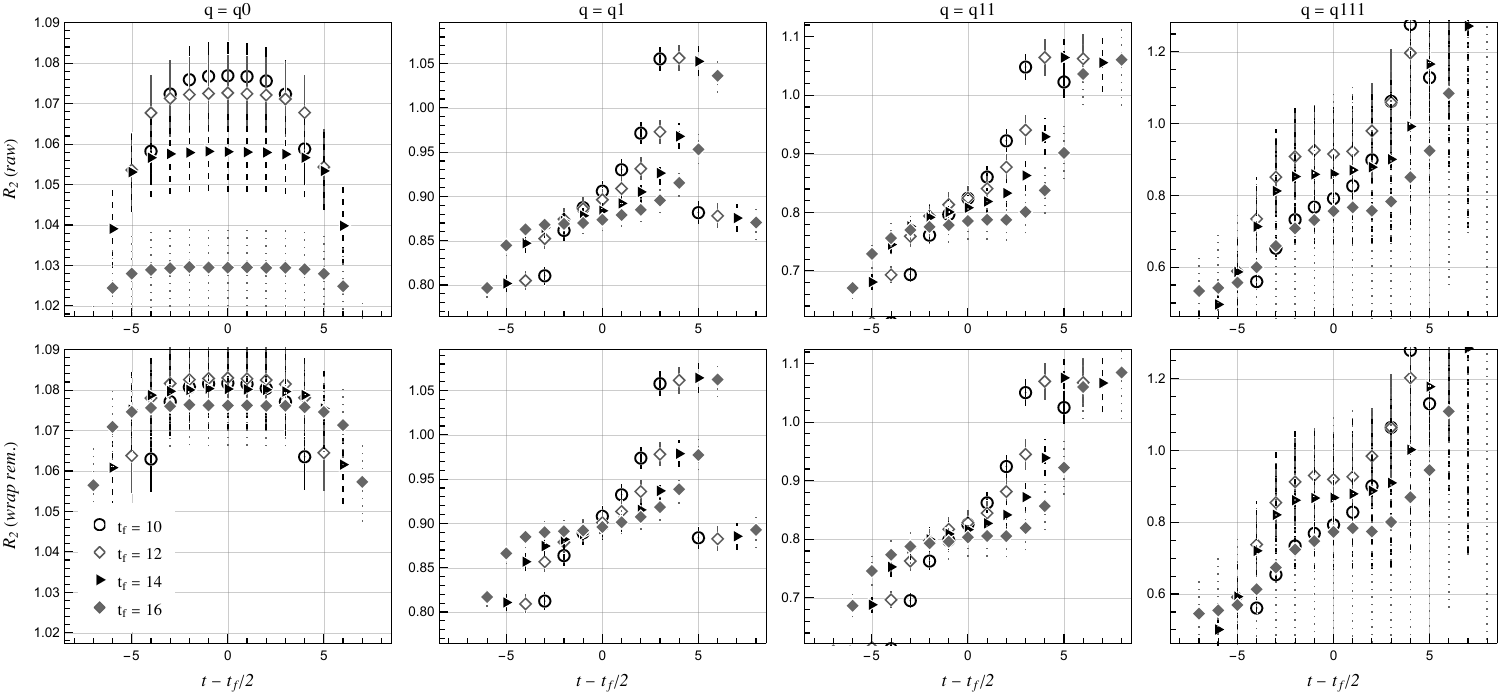}
\caption{SST: $R_2$ versus $t-t_f/2$ for all momentum shells:
$\mathbf n=(0,0,0)$, $(1,0,0)$, $(1,1,0)$, $(1,1,1)$.
Statistical errors are obtained by jackknife.
Top row is the raw ratio and bottom row is wrap-corrected. 
At zero momentum the correction largely
removes the separation dependence, whereas residual variation is visible
at nonzero momentum. The $(1,1,1)$ channel is statistically noisy.}
\label{fig:sst-r2}
\end{figure}

Fig.~\ref{fig:sst-r2} presents the conventional SST measurements.
The largest-momentum shell has a much less precise ratio, consistent
with the noisy moving-pion two-point function at the full source-sink
separation discussed in the fitted-$G_2$ analysis. The purpose of
including the full time dependence here is to make the quality of the
underlying ratio visible rather than to infer a new plateau estimate.

\begin{figure}[ht]
\centering
\includegraphics[width=0.98\textwidth,height=0.6\textheight,keepaspectratio]{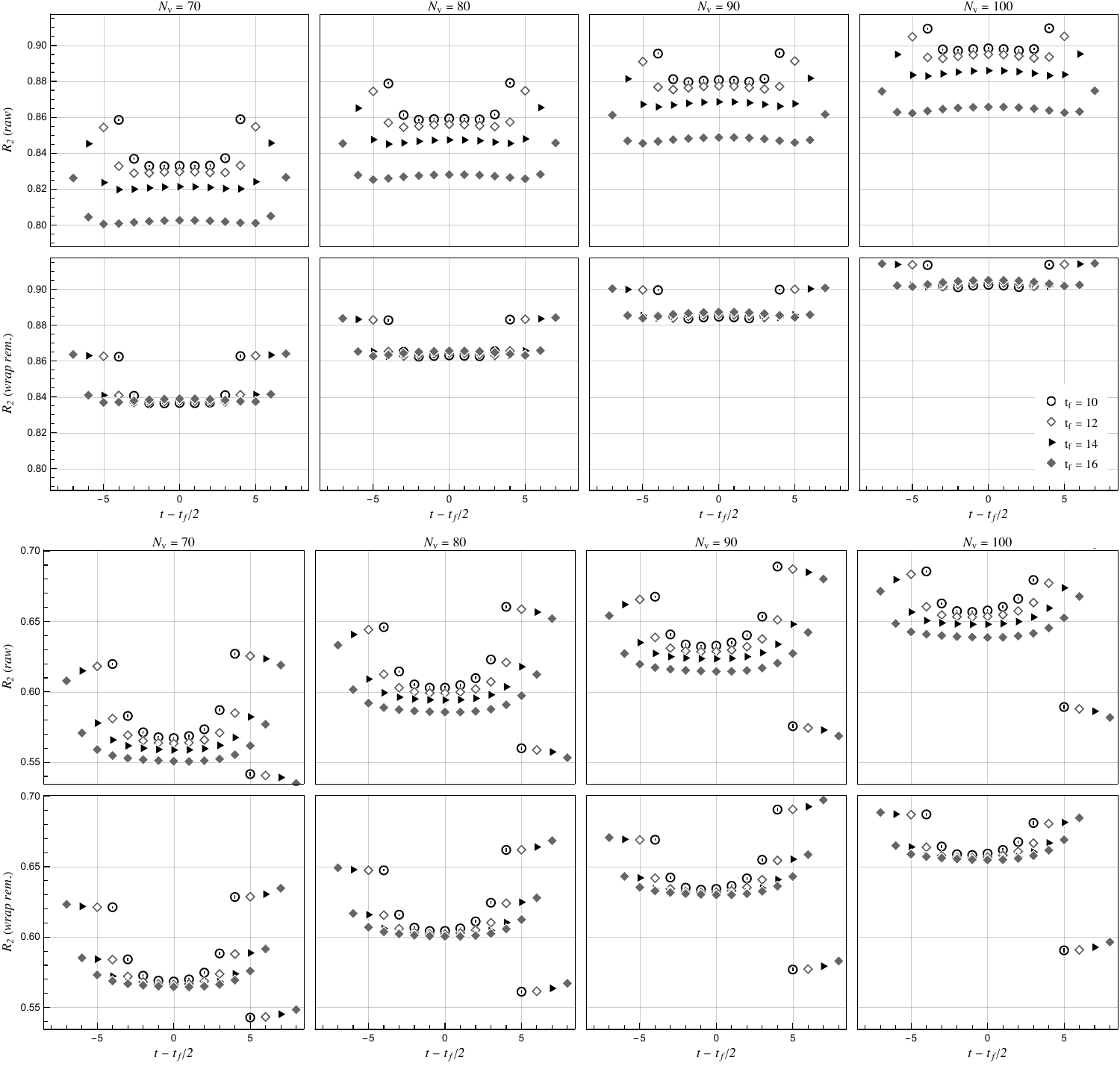}
\caption{LapH \texttt{loc}: $R_2$ versus $t-t_f/2$ with fixed scales for comparison purposes.
Top two rows: $\mathbf n=(0,0,0)$; bottom two rows: $\mathbf n=(1,0,0)$.
Columns correspond to $N_v=70,80,90,100$. Within a given momentum, the upper and lower
rows show raw and wrap-corrected ratios, respectively.
The ratios  exhibit a visible rank dependence.
}
\label{fig:loc-r2-a}
\end{figure}

\begin{figure}[ht]
\centering
\includegraphics[width=0.98\textwidth,height=0.6\textheight,keepaspectratio]{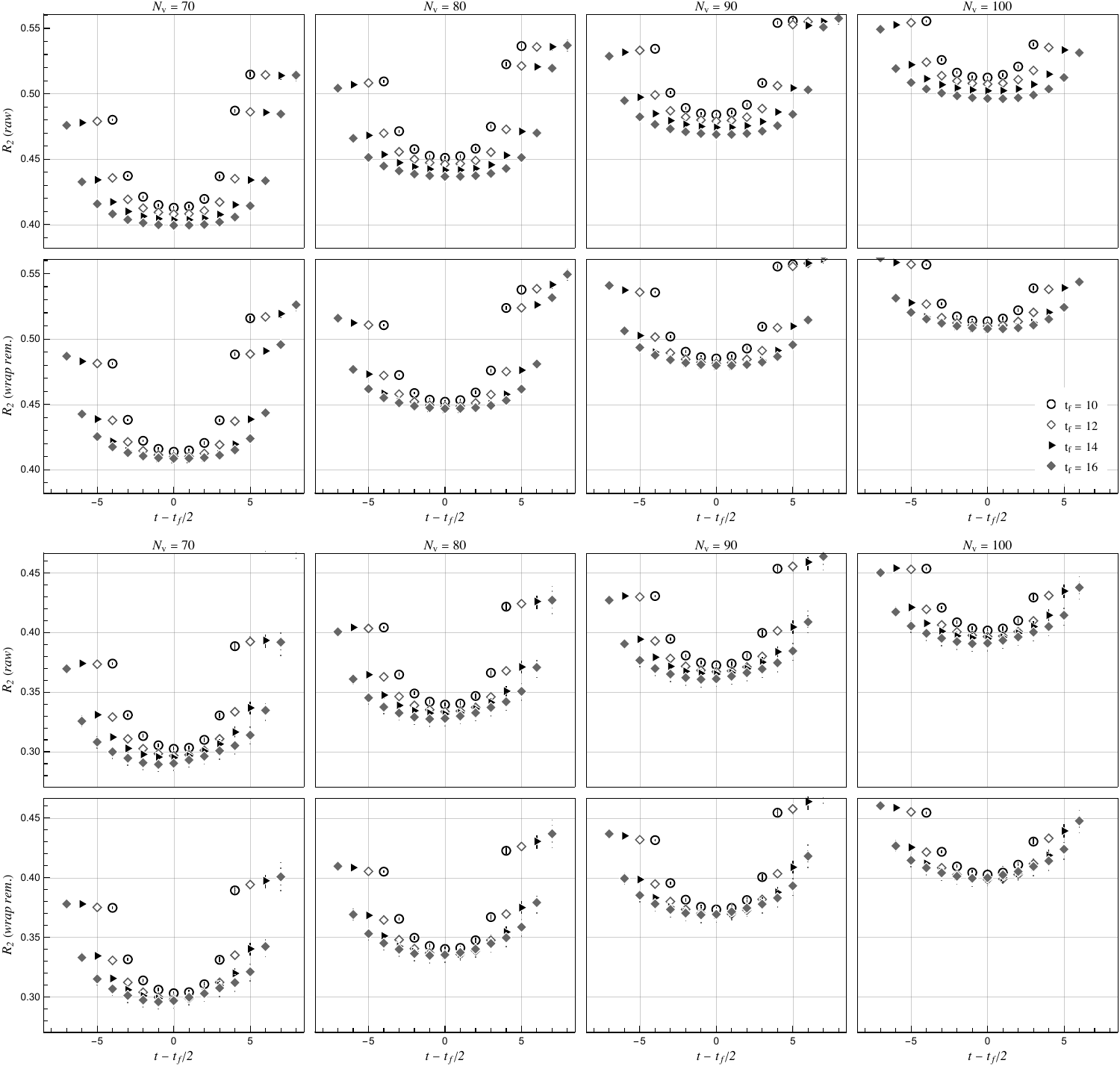}
\caption{LapH \texttt{loc}: same as Fig.~\ref{fig:loc-r2-a}, but for 
$\mathbf n=(1,1,0)$ (top two rows) and $(1,1,1)$ (bottom two rows).
}
\label{fig:loc-r2-b}
\end{figure}

For the \textbf{LapH loc} construction, the LapH projector rank is varied
simultaneously at all three operators, and the two-point correlators
entering the ratio are matched at the same rank. Fig.~\ref{fig:loc-r2-a}
and Fig.~\ref{fig:loc-r2-b} retain all four ranks. Thus the displayed rank
dependence includes changes to the interpolating operators as well as
to the current. The extrapolation in the main text is applied to the
normalized midpoint estimates, not independently to every insertion-time
point displayed.

\begin{figure}[ht]
\centering
\includegraphics[width=0.98\textwidth,height=0.6\textheight,keepaspectratio]{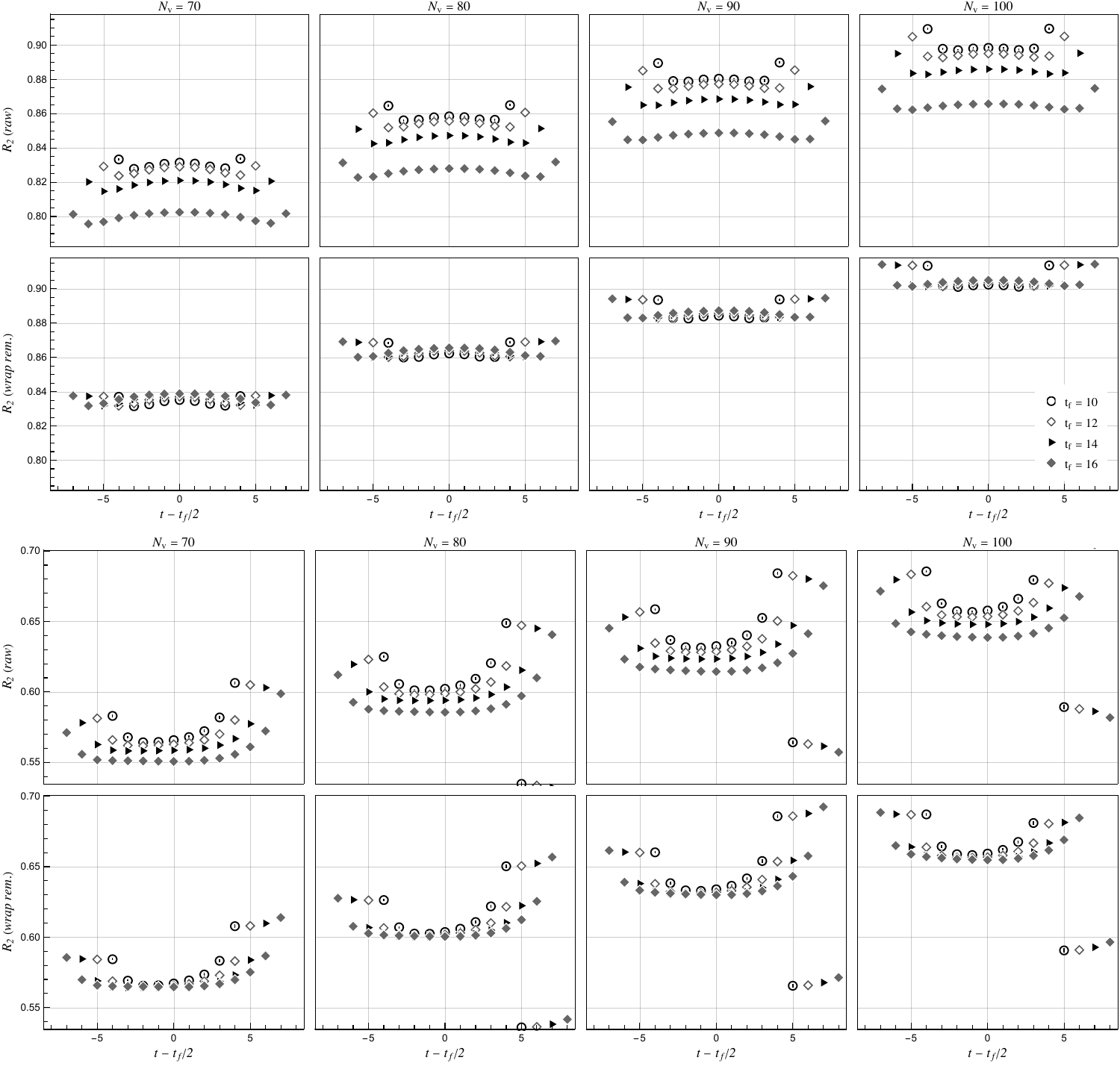}
\caption{LapH \texttt{loc\_cur}: 
$R_2$ versus $t-t_f/2$ with fixed scales for comparison purposes.
Top two rows: $\mathbf n=(0,0,0)$; bottom two rows: $(1,0,0)$.
Columns correspond to
$N_v=70,80,90,100$. Within a given momentum $\mathbf n$, the upper and lower
rows show raw and wrap-corrected ratios, respectively.
The ratios  exhibit a visible rank dependence.}
\label{fig:loccur-r2-a}
\end{figure}

\begin{figure}[ht]
\centering
\includegraphics[width=0.98\textwidth,height=0.6\textheight,keepaspectratio]{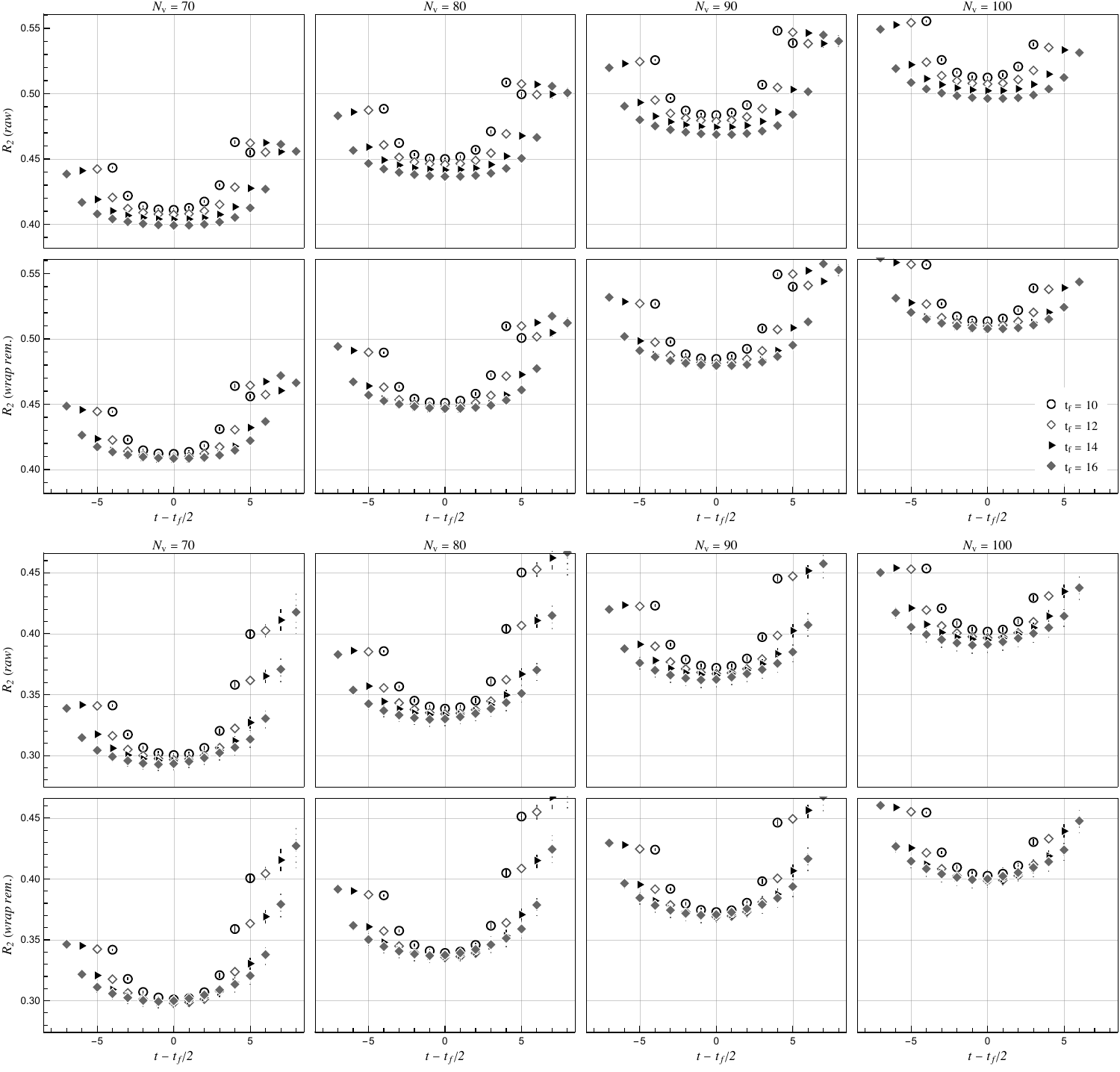}
\caption{LapH \texttt{loc\_cur}: same as Fig.~\ref{fig:loccur-r2-a}, but for 
$\mathbf n=(1,1,0)$ (top two rows), and $\mathbf n=(1,1,1)$ (bottom two rows).}
\label{fig:loccur-r2-b}
\end{figure}

For \textbf{LapH loc\_cur}, the source and sink projectors, and the
two-point functions entering the ratio, remain fixed at $N_v=100$;
only the $N_v$ for current projection is varied. This comparison isolates the
rank dependence of the current more directly than LapH \texttt{loc}, although
it does not itself establish the local-current limit or determine an
independent current-renormalization factor. Fig.~\ref{fig:loccur-r2-a}
and Fig.~\ref{fig:loccur-r2-b} provide the corresponding insertion-time data.
Their close agreement with the \texttt{LapH loc} extrapolated values is a
useful consistency check, not an estimate of all systematic errors.

\onecolumngrid

\clearpage
\twocolumngrid
\bibliography{xpion}

\end{document}